# Combine Umbrella Sampling with Integrated Tempering Method for Efficient and Accurate Calculation of Free Energy Changes of Complex Energy Surface

*Mingjun Yang,[1] Lijiang Yang,[2] Yiqin Gao,[2] and Hao Hu[1*]*

1 Department of Chemistry, The University of Hong Kong, Pokfulam Road, Hong Kong

2 College of Chemistry and Molecular Engineering, Peking University, Beijing, China

ABSTRACT

Umbrella sampling is an efficient method for the calculation of free energy changes of a system along well-defined reaction coordinates. However, when multiple parallel channels along the reaction coordinate or hidden barriers in directions perpendicular to the reaction coordinate exist, it is difficult for conventional umbrella sampling methods to generate sufficient sampling within limited simulation time. Here we propose an efficient approach to combine umbrella sampling with the integrated tempering sampling method. The umbrella sampling method is applied to conformational degrees of freedom which possess significant barriers and are chemically more relevant. The integrated tempering sampling method is employed to facilitate the sampling of other degrees of freedom in which statistically non-negligible barriers may exist. The combined method is applied to two model systems and show significantly improved sampling efficiencies as compared to standalone conventional umbrella sampling or integrated tempering sampling approaches. Therefore, the combined approach will become a very efficient method in the simulation of biomolecular processes which often involve sampling of complex rugged energy landscapes.

ABBREVIATIONS
1-D, 1-dimensional; 2-D, 2-dimensional; ITS, integrated tempering sampling; ITS-US, combined integrated tempering sampling and umbrella sampling; MD, molecular dynamics; PMF, potential of mean force; RC, reaction coordinate; US, umbrella sampling; WHAM, weighted histogram analysis method

1. Introduction

Efficient and accurate calculation of free energy change along a well-defined reaction coordinate (RC) is a central question in the physical chemistry of many important chemical and biomolecular processes. Many simulation methods have been developed to calculate the free energy and its change along the RC.[1-8] Among them, umbrella sampling (US) is one of the most widely employed methods.[9] There are two key ingredients in the US method: one is the identification of one or multiple reaction coordinates, and the other is the design of biasing potentials as a function of the RC.

Given the complexity of the energy landscape of multi-atom systems in condensed phases, the identification of the RC itself becomes a challenging issue in theoretical chemistry.[10-11] In typical US simulations, RC is defined as a combination of simple geometric terms, such as bond lengths, bond angles, and dihedral angles from chemical intuitions. It is thought that the combination of the small number of selected geometric properties can accurately characterize the progress of the reaction process. In simple reaction systems this type of definition is often sufficient for capturing the essence of the reaction processes.

Once the RC is determined, a discrete set of RC values is selected to cover the whole range of the reaction process. For each RC value, a biasing potential is applied to simulations to generate appropriate sampling of the system in the vicinity of the given value of RC. The biasing potential is expected to change the relative Boltzmann weight of different conformations along the RC. As a result, sufficient sampling can be obtained for those conformations that originally have small statistical weights and are difficult to sample in normal simulations. Once the biasing potential is decided, US simulations can be carried out in parallel. The trajectories from all simulations can be combined together with posterior analysis methods such as the weighted histogram analysis method (WHAM)[12-15] or the maximum likelihood method.[16-18] The free energy change along the RC, or the potential of mean force (PMF) of the reaction process, can be reconstructed too.

After applying the biasing potential, the US simulation is technically identical to normal MD simulation. Therefore, it will suffer any technical difficulties that normal MD might experience. Put the issue of correctness of RC aside, the success of the US then depends on the (approximately) converged sampling in all the degrees of freedom orthogonal to the reaction coordinate in conformational space. Even though in many chemical reactions this condition is likely to be fulfilled, it would become a serious issue if there are multiple reaction channels along the reaction coordinate and the transitions between different channels are inadequately sampled in US simulations.

An example 2-D potential energy landscape for this scenario is illustrated in Fig. 1. Obviously, the correctness of the 1-D US simulations depends on the converged sampling in the vertical direction. As there are two parallel paths connecting stationary states A and B, the 1-D PMF along the RC must reflect the proper statistical weights of the two paths, in particular the correct sampling of two transition states, $C_1$ and $C_2$. If a biasing potential is applied to RC which limits the sampling of RC to regions close to where $C_1$ and $C_2$ are, the corresponding US simulation must be able to sample both $C_1$ and $C_2$ in the same trajectory. This suggests there must be sufficient events to cross the barrier of C relative to $C_1$ and $C_2$. In that sense the region C becomes a *hidden barrier* for conformational transition between $C_1$ and $C_2$ along the vertical direction. One must also be reminded that the importance of sufficient sampling of C only appears when localized biased sampling methods like US are employed. In ordinary sampling, if the length of the simulation is not a concern, direct sampling of $C_1$ and $C_2$, rather than C, is the most critical issue. The latter is determined by the free energy difference between $C_1$ and A or B, and between $C_2$ and A or B. Usually it is assumed that the free energy difference between $C_1$ and A or B, or between $C_2$ and A or B, is much larger than that between $C_1$ or $C_2$ and C. That is, the target process is still dominated by the conformational transition along the RC.

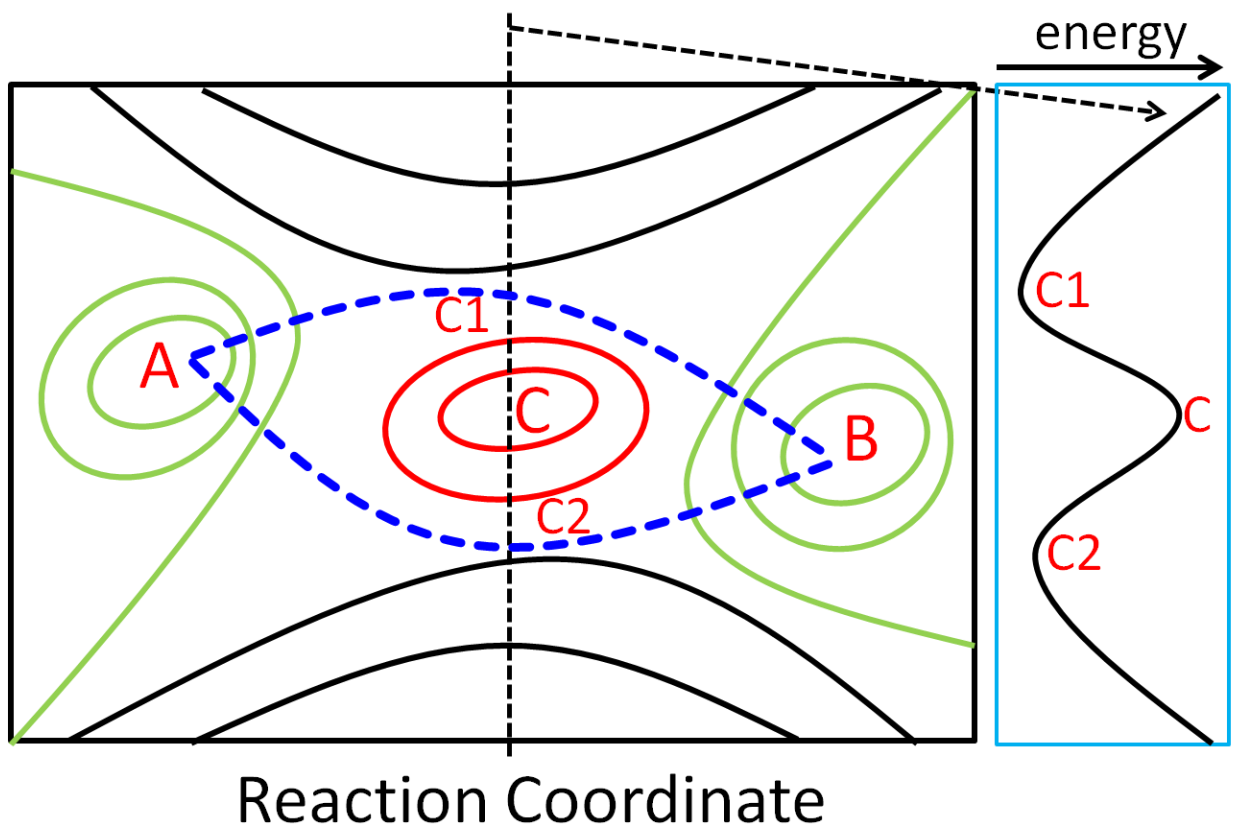

**Figure 1.** An illustrative potential energy landscape for the parallel reaction paths. A and B are two energy minima. There are two reaction paths between A and B (dashed thick blue line). X-axis is regarded as the reaction coordinate for the transition between A and B. C is a high-energy region often regarded as a hidden barrier in the direction perpendicular to the reaction coordinate. $C_1$ and $C_2$ are the transition states of the two transition paths, respectively.

Given this example, hidden barrier and parallel reaction paths appear to be two terms reflecting the same issue in conformational sampling. As shown, the existence of parallel paths along the RC already suggests a barrier in the perpendicular direction. Otherwise, the two paths would merge into one statistically significant path. For regions close to the transition states of the two paths, the correct simulation results would require converged sampling of two energy minimal regions as suggested by the energy curve of the intersecting plane along the vertical direction. If the height of the hidden barrier is low and thus can be sufficiently sampled within the length of US simulations, the 1-D US results would be correct. As the length of individual US simulation usually is significantly shorter than that of normal MD, the evolution of the system along the reaction process will likely limit to narrow regions near one reaction path if the hidden barrier is high. Therefore, 1-D US simulations could not provide correct results.

In principle one could perform multi-dimensional US for the situation illustrated here, if one knows that there are hidden barriers in directions other than the reaction coordinate. There are, however, several technical difficulties for doing so. First, the existence of hidden barrier is not always known a prior. Second, the number of independent US simulations would grow exponentially with the dimensionality of US. It is often already difficult to define properly one RC, identifying additional RC would be even more challenging. Furthermore, multi-dimensional US sampling, e.g., for a potential energy landscape depicted in Fig. 1, would unavoidably waste many simulations in the corner regions which are statistically very insignificant and unnecessary.

In addition to umbrella sampling, a large group of enhanced sampling methods has been developed which often does not require an explicit definition of reaction coordinate. Great attention has been paid in recent years to the class of generalized ensemble methods. It includes the methods of replica-exchange Monte Carlo or molecular dynamics,[19-20] simulated tempering,[21-22] multicanonical ensemble,[23-25] and the integrated tempering sampling (ITS).[26] Each method has shown distinct advantages and efficiency in tackling different problems. Of these methods, ITS is performed on an effective potential constructed by averaging over multiple

Boltzmann distributions at different temperatures. Subsequent ITS simulations can be conducted only at temperature of interests, instead of multiple copies at different temperatures. Thus ITS has relatively lower demands for computational resources. Applications of ITS have shown that the sampling of high-energy region can be remarkably enhanced[27-31] and the efficiency is at least as good as other enhanced sampling method.[27, 32] ITS needs not to use RCs, thus it can be applied to complex processes with rugged energy landscape such as protein folding.[30-31] This is one of the advantages of ITS. On the other hand, statistical weights for high-energy regions are uniformly increased regardless of their relevance to the target process, unless a small subset of degrees of freedom is selected as in the selected ITS (SITS) method.[32] This would become a disadvantage if ITS was applied to processes with practically well-known reaction coordinates, because in this case significant amount of the simulation efforts would be spent in sampling regions both statistically and chemically unimportant.

The unique features of ITS method, i.e. single-copy simulations and non-specifically enhanced sampling in phase space, make it a convenient method to combine with RC guided multiple-window methods like US in the simulation of conformational transition and reaction processes in complex systems. Here we report our development and application of a new simulation approach combining both ITS and US methods. We show that this combined approach can achieve significantly improved sampling efficiency for complex landscape like the one depicted in Fig. **1**. The theory will be provided in next section, followed by computational details of the application to two model systems. Simulation results of the conformational dynamics of butane and Ace-Nme molecules using different sampling strategies will be discussed and compared in the results section. The mechanisms of the sampling enhancement, possible extensions and applications of the new method are discussed afterwards.

2. Theory

In this section, we first outline the ITS method, which is followed by the theory of combined ITS and US method. Later the computational details for the application of the methods to two molecular systems are provided.

2.1. Integrated tempering sampling (ITS)

ITS was originally developed by Gao and co-workers.[26-27, 33-34] In this method, a generalized non-Boltzmann ensemble was constructed by summing canonical distributions over a set of different temperatures. The generalized distribution can be written as

$$P(U(\mathbf{R})) = \sum_{k=1}^{N} n_k e^{-\beta_k U(\mathbf{R})} \tag{1}$$

where $U(\mathbf{R})$ is the potential energy as a function of the coordinates $\mathbf{R}$ of the molecule and $n_k$ is a weighting factor for the temperature $T_k$ with

$$\beta_k = 1/k_B T_k \tag{2}$$

The configurational partition function of the system at a given temperature is

$$Z_k = \int e^{-\beta_k U(\mathbf{R})} d\mathbf{R} \tag{3}$$

The value of $n_k$ can be determined by applying the condition of $n_1 Z_1 = n_2 Z_2 = \cdots = n_N Z_N$, which ideally will produce a nearly flat distribution for energies sampled in the range of temperature $[T_1, T_N]$. In practice $n_k$ will have to be determined with short trial simulations prior to the production run. An effective potential energy can then be defined for the production simulation temperature $T_0$ as

$$e^{-\beta_0 \tilde{U}(\mathbf{R})} = P(U(\mathbf{R})) = \sum_{k=1}^{N} n_k e^{-\beta_k U(\mathbf{R})} \tag{4}$$

or

$$\tilde{U}(\mathbf{R}) = \frac{-1}{\beta_0} \ln \sum_{k=1}^{N} n_k e^{-\beta_k U(\mathbf{R})} \qquad (5)$$

According to this equation, one practical advantage of the ITS scheme is that the value of $\tilde{U}(\mathbf{R})$ and $U(\mathbf{R})$ can be uniquely mapped once the coefficients $\{n_k\}$ are known. The corresponding force for propagating coordinates in MD simulations can thus be computed with

$$\tilde{\mathbf{F}}_i = -\frac{\partial \tilde{U}(\mathbf{R})}{\partial \mathbf{R}_i} = \frac{\sum_{k=1}^{N} n_k \beta_k e^{-\beta_k U(\mathbf{R})}}{\beta_0 \sum_{k=1}^{N} n_k e^{-\beta_k U(\mathbf{R})}} \mathbf{F}_i \qquad (6)$$

where

$$\mathbf{F}_i = -\frac{\partial U(\mathbf{R})}{\partial \mathbf{R}_i} \qquad (7)$$

is the force of the original potential energy of the system[26, 33]. As a result, the sampling by ITS simulation can be interpreted as an averaged samplings from normal MD simulations each at temperature $T_k$ and with the potential energy $U(\mathbf{R})$, and with a weighting factor of

$$p_k = \frac{\int n_k e^{-\beta_k U(\mathbf{R})} d\mathbf{R}}{\int e^{-\beta_0 \tilde{U}(\mathbf{R})} d\mathbf{R}} = \frac{n_k Z_k}{\sum_{j=1}^{N} n_j Z_j} \qquad (8)$$

It is obvious from Eq. 6 that this method is easy to implement into current simulation packages since only original potential energies and forces are required.

As described here, the sampling enhancement of ITS can be attributed to two factors. Firstly, the ITS MD simulation under effective potential $\tilde{U}(\mathbf{R})$ at $\beta_0$ includes contribution of samples from original potential $U(\mathbf{R})$ at higher temperature $\beta_k (T_k > T_0)$ as seen from Eq. 8. This factor ensures that the ITS simulation can overcome high-energy barriers more efficiently than that of normal MD simulations under $U(\mathbf{R})$ and $\beta_0$. Secondly, the low- and high-energy regions of effective potential $\tilde{U}(\mathbf{R})$ coincide with these of the original potential $U(\mathbf{R})$ since $\tilde{U}(\mathbf{R})$ is a monotonic function of $U(\mathbf{R})$ from Eq. 5. This implies that the low-energy regions of $\tilde{U}(\mathbf{R})$ accessible more frequently in ITS simulations also corresponds to the relative low-energy region in $U(\mathbf{R})$. This feature help to maintain good balance of samples between low- and high-energy regions and thus can produce proper reweighting from $\tilde{U}(\mathbf{R})$ to $U(\mathbf{R})$ in calculation of the unbiased canonical distribution $\rho(\mathbf{R})$. By taking the two factors together, ITS simulation is equivalent to perform a trajectory on a modified energy surface with lower barriers in comparison to that of normal MD simulation on original potential, which can result in much efficient barrier transition of the system.

2.2. Combining ITS with US

In typical umbrella sampling simulations,[14-15, 35-36] the target molecular event is characterized by a preselected reaction coordinate which is thought to best characterize the reaction progress. A biasing potential, often in the form of a quadratic function,

$$V_i^b(\Omega(\mathbf{R})) = \frac{1}{2} K_i (\Omega(\mathbf{R}) - \omega_i)^2 \qquad (9)$$

is applied to each different parallel simulations. Here $K_i$ and $\omega_i$ are preset force constant and center of the distribution, $\Omega(\mathbf{R})$ is the reaction coordinate as a function of atomic positions. The potential energy in each MD simulation of US, often termed as a "sampling window", is

$$U_i(\mathbf{R}) = U(\mathbf{R}) + V_i^b(\mathbf{R}) \qquad (10)$$

By varying $\omega_i$ and correspondingly $K_i$, one can force the system to sample through the whole conformational space of the reaction coordinate $\Omega(\mathbf{R})$, even for regions with high free energy and hardly being observed in normal MD with unbiased energy function.

As proposed in the introduction, enhanced sampling technique such as ITS can be combined with umbrella sampling to improve the sampling in the degrees of freedom orthogonal to RC. When ITS is used, the biased total potential energy for the $i^{th}$ window should be $\tilde{U}_i^b(\mathbf{R})$,

$$\tilde{U}_i^b(\mathbf{R}) = \tilde{U}(\mathbf{R}) + V_i^b(\Omega(\mathbf{R})) \tag{11}$$

where $\tilde{U}(\mathbf{R})$ is the effective ITS potential defined in Eq. 5. Here we can assume that the same effective ITS potential is applied to different US sampling windows. But more generally, one can define the biased potential energy of each US window as

$$\tilde{U}_i^b(\mathbf{R}) = \tilde{U}_i(\mathbf{R}) + V_i^b(\Omega(\mathbf{R})) \tag{12}$$

where $\tilde{U}_i(\mathbf{R})$ is the effective ITS potential separately determined for each sampling window $i$. Even though this general scheme likely would be more efficient in sampling, the current work adopted the single ITS potential scheme defined in Eq. 11. The equations below still assume the general case though.

By employing the biased potential defined in Eq. 12, the unbiased canonical distribution $\rho_i(\mathbf{R})$ of the original potential energy can be recovered from the biased distribution $\rho_i^b(\mathbf{R})$ through,

$$\rho_i(\mathbf{R}) = \rho_i^b(\mathbf{R}) e^{\beta_0(\tilde{U}_i(\mathbf{R}) - U(\mathbf{R}) + V_i^b(\Omega(\mathbf{R})))} e^{-\beta_0 f_i} \tag{13}$$

where the term $f_i$ accounting for the free energy correction to the sampling window can be computed with

$$f_i = -\frac{1}{\beta_0}\ln\frac{Z_i^b}{Z_i} = -\frac{1}{\beta_0}\ln\int e^{-\beta_0\left(\tilde{U}_i(\mathbf{R})-U(\mathbf{R})+V_i^b(\Omega(\mathbf{R}))\right)}\rho_i(\mathbf{R})d\mathbf{R}. \tag{14}$$

Here $Z_i^b$ and $Z_i$ are the canonical partition function for the biased and unbiased system, respectively.

WHAM can be used to analyze the simulation data if one treats the combined term $\tilde{U}_i(\mathbf{R})-U(\mathbf{R})+V_i^b(\Omega(\mathbf{R}))$ as the biasing potential. Following the original derivation[15], the values of $\{f_i\}$ can be determined iteratively through

$$\rho(\mathbf{R}) = \sum_{i=1}^{N_W}\frac{m_i\rho_i^b(\mathbf{R})}{\sum_{j=1}^{N_W}m_j e^{-\beta_0\left(\left(V_j^b(\mathbf{R})-f_j\right)+\left(\tilde{U}_j(\mathbf{R})-U(\mathbf{R})\right)\right)}} \tag{15}$$

and

$$e^{-\beta_0 f_k} = \int e^{-\beta_0\left(\tilde{U}_k(\mathbf{R})-U(\mathbf{R})+V_k^b(\mathbf{R})\right)}\rho(\mathbf{R})d\mathbf{R}$$
$$= \sum_{i=1}^{N_W}\sum_{l=1}^{m_i}\frac{e^{-\beta_0\left(\tilde{U}_k(\mathbf{R}_{i,l})-U(\mathbf{R}_{i,l})+V_k^b(\mathbf{R}_{i,l})\right)}}{\sum_{j=1}^{N_W}m_j e^{-\beta_0\left(\left(V_j^b(\mathbf{R}_{i,l})-f_j\right)+\left(\tilde{U}_j(\mathbf{R}_{i,l})-U(\mathbf{R}_{i,l})\right)\right)}}. \tag{16}$$

Here $m_i$ is the number of samples recorded in the *i*-th simulation window.

After converged values of $\{f_i\}$ have been obtained, the unbiased probability distribution $\rho(\mathbf{R})$ can be computed from Eq. 15. Then distribution of other quantities can then be derived from $\rho(\mathbf{R})$, e.g., the reweighted probability distribution along a collective variable $\Theta(\mathbf{R})$,

$$\rho(\theta) = \int \rho(\mathbf{R})\delta(\Theta(\mathbf{R})-\theta)d\mathbf{R} \tag{17}$$

which can be conveniently evaluated from the recorded snapshots. The associated PMF along $\Theta(\mathbf{R})$ can be computed as

$$A(\theta) = -\frac{1}{\beta_0} \ln \rho(\theta) \tag{18}$$

2.3. Computational details

The ITS method and the combined ITS-US method were implemented in an in-house program QM4D.[37] The methods were applied to the simulation of two systems, the butane molecule and the Ace-Nme molecule, both in gas phase (Fig. **2**). In all simulations, MD time step was 1 fs. The temperature of the simulation system was maintained at 300 K by Langevin dynamics.[38] $\{n_k\}$ values in ITS and ITS-US simulations were determined following the optimization procedure suggested by Gao.[33]

The butane molecule was treated by the classical CHARMM force field.[39] Several different sampling strategies, including normal MD, ITS, 1-D US, and the combined ITS-US methods, were carried out to examine the potential of mean force along the rotation of the C1-C2-C3-C4 dihedral angle (Fig. **2**A). Eight independent simulations, each of 32 ns, were carried out for normal MD sampling. For US and ITS-US, 40 evenly distributed sampling windows for the dihedral angle C1-C2-C3-C4 in the range of (-180°, 180°] were selected. In each window, MD simulations of 640 ps were carried out. The harmonic restraining potential was used with a force constant of 45.0 kcal/mol/rad² at each window. For ITS, 60 different temperatures in the range of 273 ~ 450 K were used. Eight production runs were carried out with 32 ns each. The same set of ITS parameters was used in both ITS and ITS-US simulations.

The Ace-Nme molecule in gas phase was used to investigate the isomerization of the peptide bond. This molecule was treated by SCCDFTB method which is necessary for providing quantum mechanical description for the peptide bond isomerization.[40-41] The peptide bond isomerization is often assumed to proceed mainly through the rotation of the dihedral ω(CAT-CT-NZ-CAZ). However, the configuration of the nitrogen atom may become non-planar during the isomerization process (Fig. **2**B). The improper dihedral η(CT-HNZ-NZ-CAZ) can thus adopt two different states and leads to two parallel isomerization paths along ω. To explore this issue, several simulations with different sampling strategies were carried out, including 1-D US along ω, 1-D ITS-US along ω, and 2-D US along ω and η. For 2-D US, 600 ps simulations were performed for each window. Simulation time of each 1-D US window is 6000 ps. Simulation time of each 1-D ITS-US window is 1500 ps. For the dihedral ω, 54 unevenly distributed windows within (-180°, 180°] were selected with force constants of 70~160 kcal/mol/rad$^2$ used (supporting Table S1). In this case the completed cycle of *cis-trans* isomerization was simulated. For the improper dihedral η, 19 unevenly distributed windows in the range of [99°, 180°] and [-180°, -99°] were employed with a force constant of 90 kcal/mol/rad$^2$. In order to compare the sampling efficiency, the same set of restraining forces and window locations along ω was employed in 2-D US, 1-D US, and 1-D ITS-US simulations. In ITS-US simulations, 60 intermediate temperatures in the range of 273 K to 700 K were used.

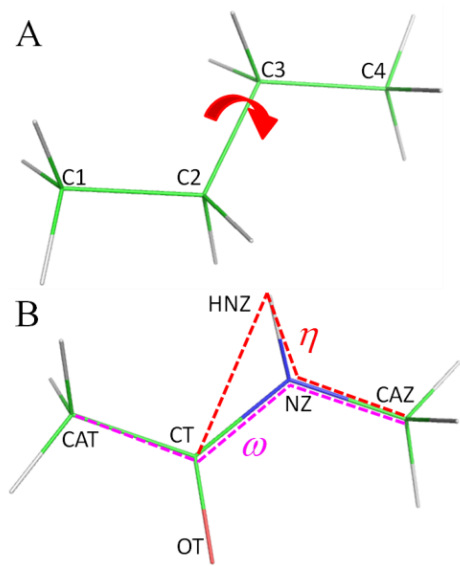

**Figure 2.** Structural model of (A) butane and (B) Ace-Nme. The two dihedrals are illustrated as ω: CAT-CT-NZ-CAZ and η: CT-HNZ-NZ-CAZ.

3. Results

3.1. Internal rotation of butane

For the butane molecule in gas phase, the conformational dynamics investigated here is the rotation of the dihedral C1-C2-C3-C4. As shown, the barrier height is ~ 5.0 kcal/mol. A barrier of this height can be well sampled in MD simulations at room temperature with 8 × 32 ns. All methods tested gave results in excellent agreement with each other, suggesting the correctness of the ITS-US method (Fig. **3**).

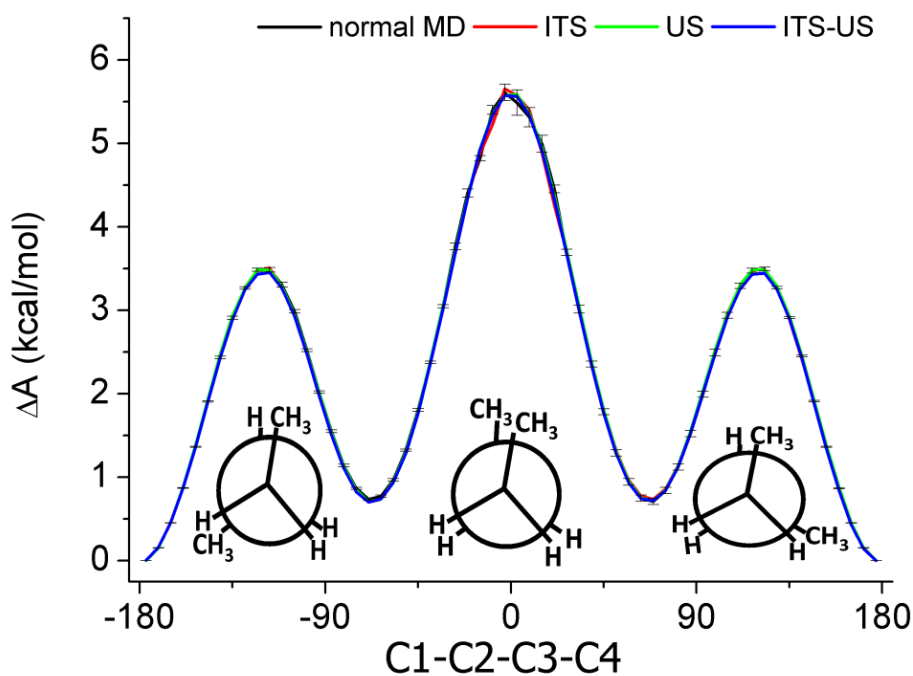

**Figure 3.** PMF profile for the rotation of dihedral C1-C2-C3-C4 in butane. The error bars were computed from eight MD trajectories of 32 ns each.

### 3.2. Peptide bond cis-trans isomerization in Ace-Nme

The Ace-Nme molecule contains a peptide bond which is the fundamental linkage between amino acids in proteins. Therefore, *cis-trans* isomerization of the peptide bond in Ace-Nme could serve a model system to provide useful information for the peptide bond isomerization in proteins and polypeptides. Obviously, one straightforward reaction coordinate for the isomerization is the dihedral ω(CAT-CT-NZ-CAZ), while a second *improper* dihedral η(CT-HNZ-NZ-CAZ) defines the chiral configuration of the nitrogen atom which is also important in the isomerization process.

We first compare the 1-D PMF along ω from different sampling strategies. The 1-D PMF results of 2-D US simulations were generated from Eq. 17 and 18 by integrating out the η degree of freedom in the joint distribution $\rho(\omega,\eta)$. Because of the extensive sampling in 2-D US simulations, we regard this result to be valid and use it as reference for comparison. Compared to the 2-D US simulations, 1-D US simulations show diverged results of 1-D PMF for simulations of different lengths (Fig. **4**A). Even for 1-D US simulations with 6000 ps per window, the results did not converge to the correct results from 2-D US simulations. On the contrary, the results of 1-D ITS-US show excellent agreement to the 2-D US results, even for simulations with only 150 ps per window (Fig. **4**B).

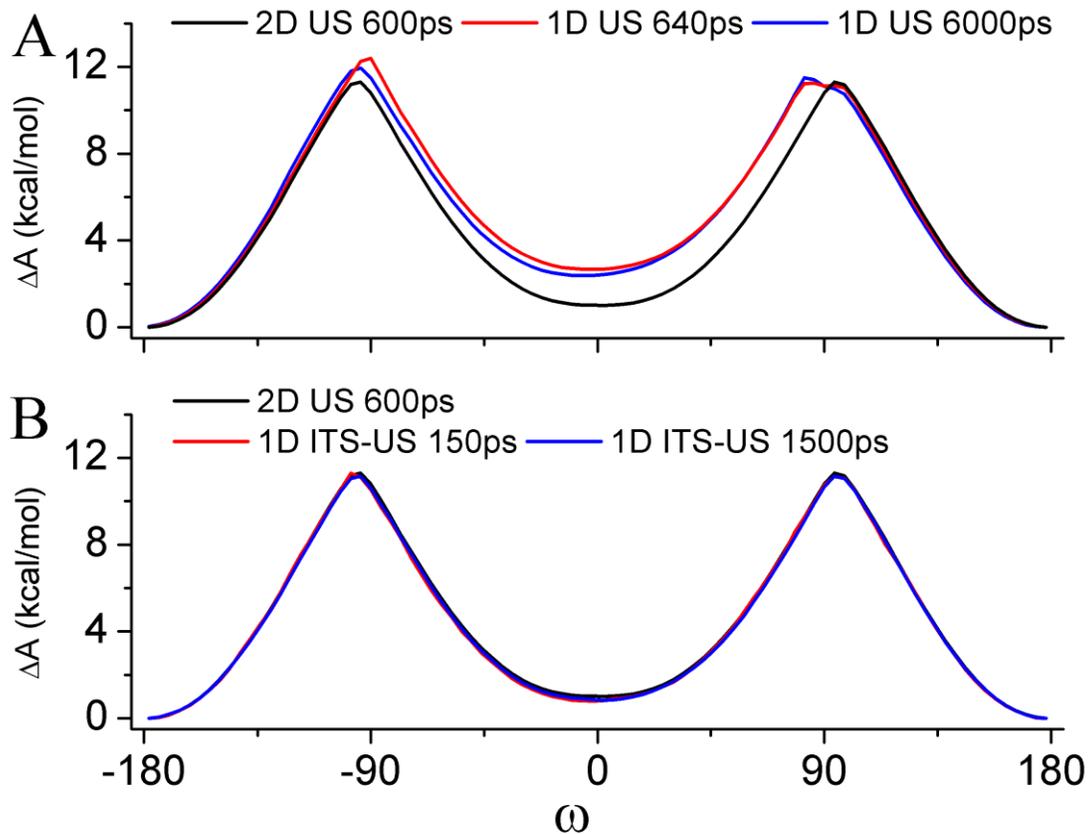

**Figure 4.** 1-D PMF profile along ω in Ace-Nme. (A) Comparison of results of conventional 1-D US and 2-D US; (B) Comparison of results of 1-D ITS-US and 2-D US.

To determine the reason why 1-D US simulations failed to provide correct PMF for the assumed isomerization process along ω, 2-D PMF was generated for each different set of simulations. It is evident that depending on the value of improper dihedral η (or the equivalent η′), isomerization along ω can proceed with two parallel paths through the transition states around η≈±130° (η′≈±50°) in either clockwise (ω: 0° to 180°) or anti-clockwise (ω: 0° to -180°) direction (Fig. **5**A). For the correct calculation of 1-D PMF along ω, the degree of freedom of η must be sufficiently sampled to reflect the correct statistical weights for conformations along the two paths. Once there are non-negligible barriers in the motions along η, conventional 1-D US sampling would not be able to efficiently cross barrier to provide converged sampling. This problem was clearly demonstrated by the 2-D PMF generated from the trajectories of 1-D US along ω (Fig. **5**B). Due to the barrier in the direction of η, the conformations sampled in different windows in 1-D US do not overlap in the transition regions along the direction of η around η≈±130° (η′≈±50°). In fact neither of the transition state of the two paths was sampled in 1-D US simulations. The poor sampling in these regions is the reason why both the height and position of the transition states in 1-D US samplings are incorrect. (Fig. 4A)

In contrast, 1-D ITS-US, even for data extracted from simulation fragments of only 150 ps, provided significantly improved samplings (Fig. **5**C and supporting Fig. S1). A significant amount of conformations in the transition region have been sampled in 150 ps of 1-D ITS-US simulations. For the parallel isomerization paths, one of the two transition states with lower free energy barrier was sampled properly, thus providing correct results for 1-D PMF of ω. When the simulation time was increased to 1500 ps per window, the 2-D PMF well reproduced the results of 2-D US illustrated by clear appearance of two parallel isomerization paths (Fig. **5**D).

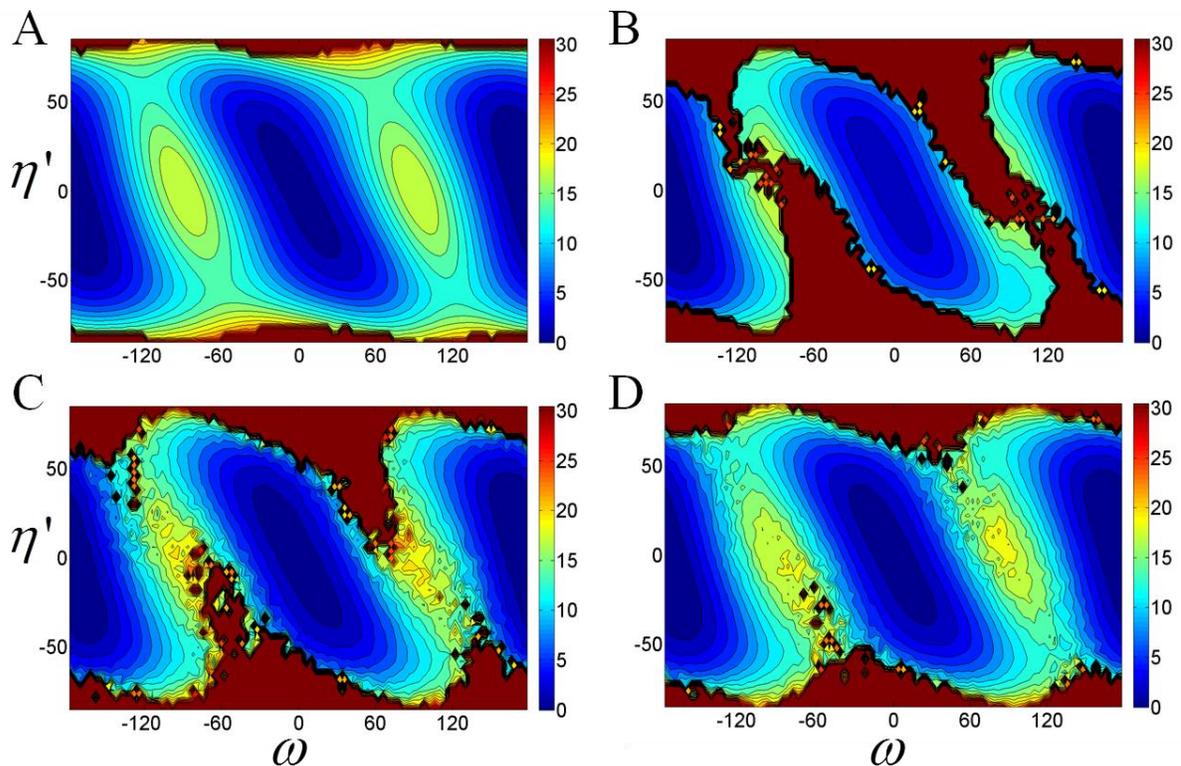

**Figure 5**. 2-D PMF maps of the Ace-Nme molecule. (A) Results reconstructed directly from 2-D US simulations of 600 ps per window; (B) Results projected from 1-D US simulations of 6000 ps per window; (C) Results projected from 1-D ITS-US simulations of 150 ps per sampling window; (D) Results projected from 1-D ITS-US simulations of 1500 ps per sampling window. Note for clarity, the Y-axis is defined as η′. η′=η-π when η>0 and η′=η+π when η<0. The unit for both axes is degree.

4. Discussions

In this study, we show a combination of ITS and US can efficiently generate satisfactory sampling for system with complex energy landscape and possibly multiple parallel transformation paths. The success of this ITS-US approach is due to the proper combination of the advantages of both methods. On one hand, the degree of freedom with significant energy barrier is more efficiently sampled by RC guided US approach. In the Ace-Nme system, the height of the barrier along ω is about 12 kcal/mol. In the current case, without RC, general enhanced sampling methods might still be able to sample the barrier but with very low efficiency. (Fig. S2 and S3, supporting information) The reason for their low efficiency is that the target process involves a very localized conformational motion. It would be the best scenario if enhanced sampling can be directly applied to the specific motions instead of spreading among all degrees of freedom.

On the other hand, with the increase of the number of degrees of freedom, energy landscape of large molecule becomes more and more complicated. Numerous minima, maxima, and transition saddle points emerge even for those degrees of freedom regarded as chemically less relevant. This rugged energy landscape is the ultimate reasons for the hierarchy of conformational motions in biomolecules. Even under the assumption that there is one motional direction in which the barrier dominates the target processes, the sampling of the directions of remaining conformational degrees of freedom with lower but non-negligible barriers remains to be important. In limited examples high-dimensional US could be employed, but with significantly increased technical difficulty and computational costs. General enhanced sampling technique, such as ITS, becomes an effective approach to sample these degrees of freedoms.

A simple comparison with the simulation time can be used to provide a rough picture for the level of improvement the ITS-US approach made with respect to normal US. For the Ace-Nme system tested using the same set of simulation parameters, 150 ps simulations of each window of 1-D ITS-US generate PMF in good agreement with reference results, while plain 1-D US simulation up to 6000 ps per window cannot provide results of same quality. The ratio for the computational cost of 1-D ITS-US and US is ~ 1:40. On the other hand, even without optimizing the number of US windows, the total simulation length of 1D ITS-US is 0.6 × 54 = 32.4 ns, which provided results comparable to or even better than 8 parallel ITS simulations with a total length of 8 × 96 = 768 ns (Fig. S2 and S3). Both ratios demonstrate again the much-improved efficiency in the ITS-US approach.

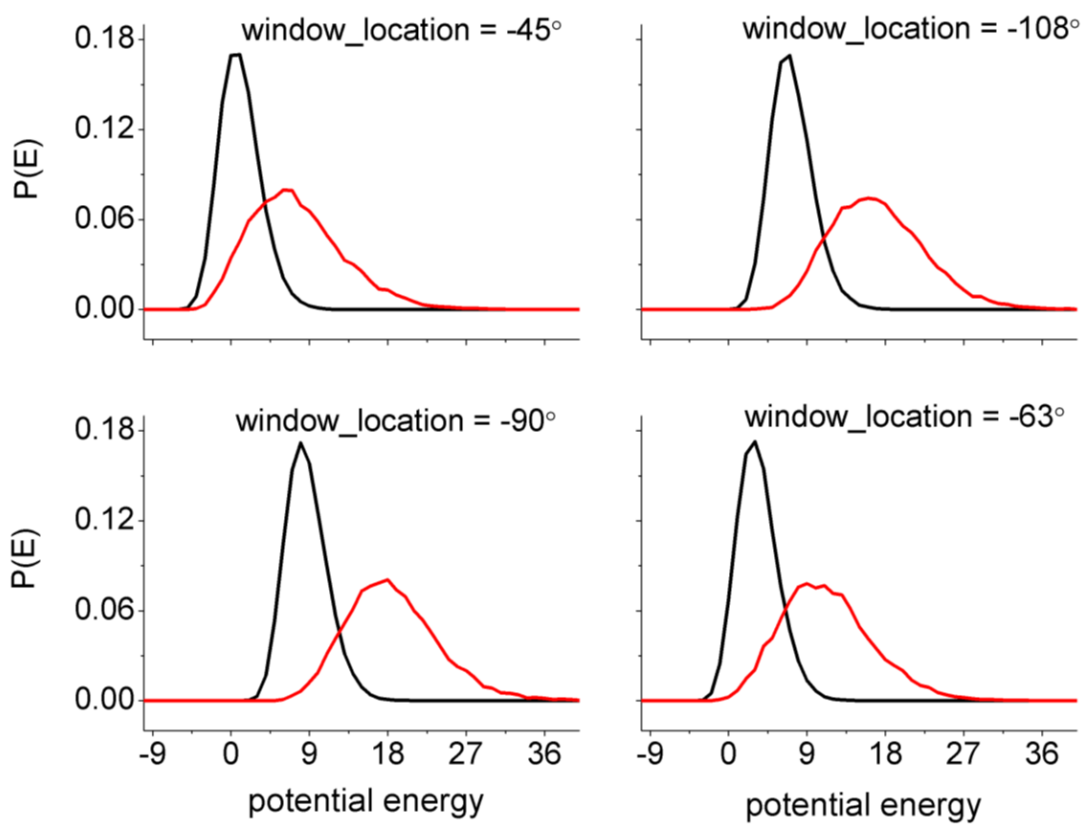

**Figure 6**. Distribution of potential energy in different simulations. 1-D US: black line; 1-D ITS-US: red line. The energy was shifted by 8453.78 kcal/mol in all panels to plot the distribution in vicinity of zero.

To show the effectiveness of the ITS-US approach, we also plotted the distribution of potential energy in different simulations. As shown, the fluctuation of potential energy in conventional US simulations is small, thus crossing a barrier of even moderate height is very difficult (Fig. **6**). In contrast, ITS samples the potential energy in a range significantly broader than normal MD, thus providing sufficient sampling for crossing barriers of low and medium heights.

We note that the current approach still possesses great potential to improve the efficiency. The first improvement could be made to use different set of $\{n_k\}$ parameters for the simulation of different US windows. Second, as shown in the theory of ITS, the relative conformational weight along RC will unavoidably be adjusted by ITS. Therefore, the restraining potential and correspondingly the number of US windows in US simulations could be re-optimized in this combined ITS-US approaches. Therefore, a smaller number of US windows might be utilized in ITS-US simulations.

Even though the current study demonstrates the high efficiency of the combined ITS-US approach, we note that this new method strongly depends on the identification of an appropriate RC. In systems where a RC is difficult to identify, e.g. protein folding, simple ITS might be practically more advantageous. Moreover, the results of ITS simulations can be analyzed to identify proper order parameters of the target process, which can be used subsequently in combined enhanced sampling simulations to improve the accuracy of the results.

Besides our efforts in improving the efficiency of US algorithm, several extensions and variants of this method have been proposed in the past several decades. Instead of performing multiple-window simulations with local biasing potential at each US window, the adaptive US method, originally proposed by Mezei[42] and Hooft et al[43] and further developed by Karplus and co-workers,[44-45] updates the biasing potential iteratively over the whole range of RC in a single simulation. The samples generated in each iteration can be combined to reconstruct PMF in posterior analysis. Similarly, some other extensions of US can also carry out the simulation in one trajectory, e.g., self-healing US, simulated tempering US, and local evaluation US. In self-healing US, a biasing potential is progressively constructed using the time-dependent probability density distribution and slow dynamics can be accelerated along the RC.[46] The local evaluation US combines the searching power of local evaluation to build up an optimized biasing potential at the first stage and the sampling ability of US to explore relevant conformational space at the second stage.[47] In this method, the biasing potential can be constructed by fragments and then used for larger molecules.[48] By combining the idea of simulated tempering with US algorithm, Mori and Okamoto make the transition attempts within parameter space of biased potential instead of the temperature space as in original simulated tempering method.[49] In such a scheme, the whole range of RCs can be sampled in one single trajectory instead of the multiple-window simulations in typical US. As an alternative way to incorporate tempering ideas in US, replica-exchange US are also proposed in several different studies, in which pairs of replicas with different parameters of biasing potentials, instead of the temperatures in original replica-exchange method, are exchanged.[50-52] In addition to geometric RCs in US extensions, Wu constructed a special biasing potential by using potential functions of two states to efficiently bridge up different phase space regions.[53] In one study of ion channels, metadynamics and US is employed in a sequential way, in which ion permeation paths were firstly identified by metadynamics and then PMF was constructed by US along the selected paths.[54] These extensions and variants of US, in addition to our newly developed ITS-US method, show different advantages and efficiency in tackling different problems, which can be employed accordingly by researchers with different purposes.

5. Conclusions

In conclusion, the ITS-US method developed here can produce remarkably improved sampling efficiency for complex energy surface such as the multiple parallel reaction paths we tested in this study. The combined approach provides great application potential for a broad range of interesting questions, such as the reaction in solution, enzymatic catalysis, and conformational transition in molecular recognition etc.

Supporting Information

For Ace-Nme system, the force constants and window locations used in US simulations in Table S1, the PMF along ω computed by 150 ps segment in Figure S1, and standalone ITS results in Figure S2 and S3. This material is available free of charge via the Internet at http://pubs.acs.org.

AUTHOR INFORMATION

**Corresponding Author**

**\*Email: haohu@hku.hk**

ACKNOWLEDGMENT

We thank the Research Grants Council of Hong Kong, the University Development Fund on Fast Algorithms, Strategic Theme on Computational Sciences, and Seed Funding for Basic Research at the University of Hong Kong for providing financial supports, the high-performance computing facility of the computer center at HKU for providing computing resources.

REFERENCES

1.	Hu, Y.; Hong, W.; Shi, Y.; Liu, H., Temperature-Accelerated Sampling and Amplified Collective Motion with Adiabatic Reweighting to Obtain Canonical Distributions and Ensemble Averages. *J. Chem. Theory Comput.* **2012,** *8* (10), 3777-3792.
2.	Zheng, L.; Chen, M.; Yang, W., Random Walk in Orthogonal Space to Achieve Efficient Free-Energy Simulation of Complex Systems. *Proc. Natl. Acad. Sci. USA* **2008,** *105* (51), 20227-20232.
3.	Hu, H.; Yang, W., Free Energies of Chemical Reactions in Solution and in Enzymes with ab initio Quantum Mechanics/Molecular Mechanics Methods. In *Annu. Rev. Phys. Chem.*, 2008; Vol. 59, pp 573-601.
4.	Darve, E.; Rodriguez-Gomez, D.; Pohorille, A., Adaptive Biasing Force Method for Scalar and Vector Free Energy Calculations. *J. Chem. Phys.* **2008,** *128* (14), 144120.
5.	In *Free Energy Calculations: Theory and Applications in Chemistry and Biology*, Chipot, C.; Pohorille, A., Eds. Springer: 2007; pp 1-511.
6.	Micheletti, C.; Laio, A.; Parrinello, M., Reconstructing the Density of States by History-Dependent Metadynamics. *Phys. Rev. Lett.* **2004,** *92* (17), 170601.
7.	Hu, H.; Yun, R. H.; Hermans, J., Reversibility of Free Energy Simulations: Slow Growth May Have a Unique Advantage. (with a Note on Use of Ewald Summation). *Mol. Simulation.* **2002,** *28* (1-2), 67-80.
8.	Hermans, J., Simple Analysis of Noise and Hysteresis in (Slow-Growth) Free-Energy Simulations. *J. Phys. Chem.* **1991,** *95* (23), 9029-9032.
9.	Kaestner, J., Umbrella Sampling. *Wiley Interdisciplinary Reviews-Computational Molecular Science* **2011,** *1* (6), 932-942.
10.	Singharoy, A.; Cheluvaraja, S.; Ortoleva, P., Order Parameters for Macromolecules: Application to Multiscale Simulation. *J. Chem. Phys.* **2011,** *134* (4), 044104.
11.	Bolhuis, P. G.; Dellago, C.; Chandler, D., Reaction Coordinates of Biomolecular Isomerization. *Proc. Natl. Acad. Sci. USA* **2000,** *97* (11), 5877-5882.
12.	Ferrenberg, A. M.; Swendsen, R. H., Optimized Monte-Carlo Data-Analysis. *Phys. Rev. Lett.* **1989,** *63* (12), 1195-1198.
13.	Kumar, S.; Bouzida, D.; Swendsen, R. H.; Kollman, P. A.; Rosenberg, J. M., The Weighted Histogram Analysis Method for Free-Energy Calculations on Biomolecules. 1. The Method. *J. Comput. Chem.* **1992,** *13* (8), 1011-1021.
14.	Roux, B., The Calculation of the Potential of Mean Force Using Computer-Simulations. *Comput. Phys. Commun.* **1995,** *91* (1-3), 275-282.
15.	Souaille, M.; Roux, B., Extension to the Weighted Histogram Analysis Method: Combining Umbrella Sampling with Free Energy Calculations. *Comput. Phys. Commun.* **2001,** *135* (1), 40-57.
16.	Lee, T. S.; Radak, B. K.; Pabis, A.; York, D. M., A New Maximum Likelihood Approach for Free Energy Profile Construction from Molecular Simulations. *J. Chem. Theory Comput.* **2013,** *9* (1), 153-164.
17.	Tan, Z. Q., On a Likelihood Approach for Monte Carlo Integration. *J. Am. Stat. Assoc.* **2004,** *99* (468), 1027-1036.
18.	Bartels, C., Analyzing Biased Monte Carlo and Molecular Dynamics Simulations. *Chem. Phys. Lett.* **2000,** *331* (5-6), 446-454.

19. Hansmann, U. H. E., Parallel Tempering Algorithm for Conformational Studies of Biological Molecules. *Chem. Phys. Lett.* **1997,** *281* (1-3), 140-150.
20. Sugita, Y.; Okamoto, Y., Replica-Exchange Molecular Dynamics Method for Protein Folding. *Chem. Phys. Lett.* **1999,** *314* (1-2), 141-151.
21. Lyubartsev, A. P.; Martsinovski, A. A.; Shevkunov, S. V.; Vorontsovvelyaminov, P. N., New Approach to Monte-Carlo Calculation of the Free-Energy - Method of Expanded Ensembles. *J. Chem. Phys.* **1992,** *96* (3), 1776-1783.
22. Marinari, E.; Parisi, G., Simulated Tempering - A New Monte-Carlo Scheme. *Europhys. Lett.* **1992,** *19* (6), 451-458.
23. Berg, B. A.; Neuhaus, T., Multicanonical Algorithms for First Order Phase Transitions. *Phys. Lett. B* **1991,** *267* (2), 249-253.
24. Berg, B. A.; Neuhaus, T., Multicanonical Ensemble - a New Approach to Simulate 1st-Order Phase-Transitions. *Phys. Rev. Lett.* **1992,** *68* (1), 9-12.
25. Nakajima, N.; Nakamura, H.; Kidera, A., Multicanonical Ensemble Generated by Molecular Dynamics Simulation for Enhanced Conformational Sampling of Peptides. *J. Phys. Chem. B* **1997,** *101* (5), 817-824.
26. Gao, Y. Q., An Integrate-over-Temperature Approach for Enhanced Sampling. *J. Chem. Phys.* **2008,** *128* (6), 064105.
27. Yang, L.; Shao, Q.; Gao, Y. Q., Comparison between Integrated and Parallel Tempering Methods in Enhanced Sampling Simulations. *J. Chem. Phys.* **2009,** *130* (12), 124111.
28. Shao, Q.; Gao, Y. Q., The Relative Helix and Hydrogen Bond Stability in the B Domain of Protein A as Revealed by Integrated Tempering Sampling Molecular Dynamics Simulation. *J. Chem. Phys.* **2011,** *135* (13), 135102.
29. Shao, Q.; Yang, L.; Gao, Y. Q., Structure Change of Beta-Hairpin Induced by Turn Optimization: An Enhanced Sampling Molecular Dynamics Simulation Study. *J. Chem. Phys.* **2011,** *135* (23), 235104.
30. Shao, Q.; Shi, J.; Zhu, W., Enhanced Sampling Molecular Dynamics Simulation Captures Experimentally Suggested Intermediate and Unfolded States in the Folding Pathway of Trp-Cage Miniprotein. *J. Chem. Phys.* **2012,** *137* (12), 125103.
31. Shao, Q.; Zhu, W.; Gao, Y. Q., Robustness in Protein Folding Revealed by Thermodynamics Calculations. *J. Phys. Chem. B* **2012,** *116* (47), 13848-13856.
32. Yang, L.; Gao, Y. Q., A Selective Integrated Tempering Method. *J. Chem. Phys.* **2009,** *131* (21), 214109.
33. Gao, Y. Q., Self-Adaptive Enhanced Sampling in the Energy and Trajectory Spaces: Accelerated Thermodynamics and Kinetic Calculations. *J. Chem. Phys.* **2008,** *128* (13), 134111.
34. Gao, Y. Q.; Yang, L.; Fan, Y.; Shao, Q., Thermodynamics and Kinetics Simulations of Multi-Time-Scale Processes for Complex Systems. *Int. Rev. Phys. Chem.* **2008,** *27* (2), 201-227.
35. Torrie, G. M.; Valleau, J. P., Monte-Carlo Free-Energy Estimates Using Non-Boltzmann Sampling - Application to Subcritical Lennard-Jones Fluid. *Chem. Phys. Lett.* **1974,** *28* (4), 578-581.
36. Torrie, G. M.; Valleau, J. P., Non-Physical Sampling Distributions in Monte-Carlo Free-Energy Estimation - Umbrella Sampling. *J. Comput. Phys.* **1977,** *23* (2), 187-199.
37. *$QM^4D$: an integrated and versatile quantum mechanical/molecular mechanical simulation package (http://www.qm4d.info/).* **2013**.
38. Van Gunsteren, W. F.; Berendsen, H. J. C., A Leap-Frog Algorithm for Stochastic Dynamics. *Mol. Simulat.* **1988,** *1* (3), 173-185.

39. MacKerell, A. D.; Bashford, D.; Bellott, M.; Dunbrack, R. L.; Evanseck, J. D.; Field, M. J.; Fischer, S.; Gao, J.; Guo, H.; Ha, S.; Joseph-McCarthy, D.; Kuchnir, L.; Kuczera, K.; Lau, F. T. K.; Mattos, C.; Michnick, S.; Ngo, T.; Nguyen, D. T.; Prodhom, B.; Reiher, W. E.; Roux, B.; Schlenkrich, M.; Smith, J. C.; Stote, R.; Straub, J.; Watanabe, M.; Wiorkiewicz-Kuczera, J.; Yin, D.; Karplus, M., All-Atom Empirical Potential for Molecular Modeling and Dynamics Studies of Proteins. *J. Phys. Chem. B* **1998,** *102* (18), 3586-3616.
40. Elstner, M.; Frauenheim, T.; Kaxiras, E.; Seifert, G.; Suhai, S., A Self-Consistent Charge Density-Functional Based Tight-Binding Scheme for Large Biomolecules. *Phys. Status. Solidi. B* **2000,** *217* (1), 357-376.
41. Elstner, M.; Porezag, D.; Jungnickel, G.; Elsner, J.; Haugk, M.; Frauenheim, T.; Suhai, S.; Seifert, G., Self-Consistent-Charge Density-Functional Tight-Binding Method for Simulations of Complex Materials Properties. *Phys. Rev. B* **1998,** *58* (11), 7260-7268.
42. Mezei, M., Adaptive Umbrella Sampling - Self-Consistent Determination of the Non-Boltzmann Bias. *J. Comput. Phys.* **1987,** *68* (1), 237-248.
43. Hooft, R. W. W.; Vaneijck, B. P.; Kroon, J., An Adaptive Umbrella Sampling Procedure in Conformational-Analysis Using Molecular-Dynamics and Its Application to Glycol. *J. Chem. Phys.* **1992,** *97* (9), 6690-6694.
44. Bartels, C.; Karplus, M., Probability Distributions for Complex Systems: Adaptive Umbrella Sampling of the Potential Energy. *J. Phys. Chem. B* **1998,** *102* (5), 865-880.
45. Bartels, C.; Karplus, M., Multidimensional Adaptive Umbrella Sampling: Applications to Main Chain and Side Chain Peptide Conformations. *J. Comput. Chem.* **1997,** *18* (12), 1450-1462.
46. Marsili, S.; Barducci, A.; Chelli, R.; Procacci, P.; Schettino, V., Self-Healing Umbrella Sampling: A Non-Equilibrium Approach for Quantitative Free Energy Calculations. *J. Phys. Chem. B* **2006,** *110* (29), 14011-14013.
47. Hansen, H. S.; Huenenberger, P. H., Using the Local Elevation Method to Construct Optimized Umbrella Sampling Potentials: Calculation of the Relative Free Energies and Interconversion Barriers of Glucopyranose Ring Conformers in Water. *J. Comput. Chem.* **2010,** *31* (1), 1-23.
48. Hansen, H. S.; Daura, X.; Huenenberger, P. H., Enhanced Conformational Sampling in Molecular Dynamics Simulations of Solvated Peptides: Fragment-Based Local Elevation Umbrella Sampling. *J. Chem. Theory Comput.* **2010,** *6* (9), 2598-2621.
49. Mori, Y.; Okamoto, Y., Free-Energy Analyses of a Proton Transfer Reaction by Simulated-Tempering Umbrella Sampling and First-Principles Molecular Dynamics Simulations. *Phys. Rev. E* **2013,** *87* (2), 3301-3301.
50. Park, S.; Kim, T.; Im, W., Transmembrane Helix Assembly by Window Exchange Umbrella Sampling. *Phys. Rev. Lett.* **2012,** *108* (10), 108102.
51. Curuksu, J.; Zacharias, M., Enhanced Conformational Sampling of Nucleic Acids by a New Hamiltonian Replica Exchange Molecular Dynamics Approach. *J. Chem. Phys.* **2009,** *130* (10), 104110.
52. Sugita, Y.; Kitao, A.; Okamoto, Y., Multidimensional Replica-Exchange Method for Free-Energy Calculations. *J. Chem. Phys.* **2000,** *113* (15), 6042-6051.
53. Wu, D., An Efficient Umbrella Potential for the Accurate Calculation of Free Energies by Molecular Simulation. *J. Chem. Phys.* **2010,** *133* (4), 044115.

54. Zhang, Y.; Voth, G. A., Combined Metadynamics and Umbrella Sampling Method for the Calculation of Ion Permeation Free Energy Profiles. *J. Chem. Theory Comput.* **2011,** *7* (7), 2277-2283.